\documentstyle[12pt]{article}

\begin{document}
\baselineskip 0.8cm
\setlength{\textwidth}{14cm}
\setlength{\textheight}{21cm}
\setlength{\topmargin}{0cm}
\setlength{\parindent}{1cm}

\newcommand{\gsim}
{ \mbox{ \raisebox{-1ex}{ $\stackrel{>}{\sim}$ } } }
\newcommand{\lsim}
{ \mbox{ \raisebox{-1ex}{ $\stackrel{<}{\sim}$ } } }

\begin{titlepage}
\begin{flushright}
TU-409
\\
July, 1992
\end{flushright}

\vskip 1.0cm
\begin{center}
{\Large \bf
Coannihilation Effects and Relic   Abundance of Higgsino-Dominant LSPs}
\vskip 1.0cm
Satoshi Mizuta
\vskip 0.7cm

{\it Department of Physics, Tohoku University, Sendai 980, Japan}
\vskip 0.7cm
and
\vskip 0.7cm

Masahiro Yamaguchi
\vskip 0.7cm

{\it Department of Physics, College of General Education
\\ Tohoku University, Sendai 980, Japan}
\vskip 0.8cm
\abstract{We calculate the relic abundance of Higgsino-dominant
lightest superparticles,
taking account of coannihilations with the
superparticles which are almost degenerate with the lightest one. We show
that their relic abundance is reduced  drastically by the coannihilation
processes and hence they are cosmologically of no interest.
}

\end{center}
\end{titlepage}

The lightest superparticle (LSP) in supersymmetric (SUSY) standard models,
mostly taken to be a neutralino, is an intriguing candidate for dark matter
\cite{Goldberg,Krauss,EHNOS}.
An important task is therefore to compute the relic abundance of the
LSPs, which is conveniently
 represented by $\Omega_{\chi} h^2$ where $\Omega_{\chi}$ is the ratio
of the mass density of the LSPs
to the critical one to close the Universe and $h$
$(0.4 \lsim h \lsim 1)$ stands for
the Hubble constant in  units of 100 km sec$^{-1}$Mpc$^{-1}$.

This issue has been discussed extensively in the literature
\cite{Griest,GKT,OliveSrednicki,MOS}, assuming (1) the minimal
particle content of the SUSY standard model, (2) a GUT relation on gaugino
mass parameters (see below), (3) the neutralino LSP and (4) R-parity
conservation  which guarantees the stability of the LSP.
The abundance of the neutralino LSPs crucially depends not only
on their  mass but also on the
composition of them, since the LSP is a linear combination of neutral
gauginos, $\tilde B$, $\tilde W_3$, and
neutral Higgsinos $\tilde H_1$, $\tilde H_2$,  and each of them has different
pair annihilation processes which determine its abundance.

The results obtained previously are summarized as follows:
let us first consider the case where  the LSP is lighter than the W-boson.
If it is almost a pure gaugino  or a pure
Higgsino, its  density can  reach (or even  exceed) the critical one
 and hence it is a good candidate for the dark
matter of the Universe. If the LSP is an admixture of the gauginos and
the  Higgsinos,
 the relic abundance is generally small. On the contrary, in the case where
the LSP is heavier than the W-boson,
 the annihilation mode to  a W-pair  opens
if it is a pure Higgsino or a mixed LSP and the relic abundance is very
small until  the mass increases to a TeV region where  $\Omega_{\chi} h^2$ of
the LSPs  becomes again of order unity. If the LSP is a gaugino,
there is no annihilation process to the W-pair and
hence it is  still  abundant in  the Universe.
If its  mass exceeds $\sim 500$ (GeV), it will
overclose the Universe and  such a parameter region is excluded.

The previous calculations have, however,
 included only  annihilations of the LSP pairs.
Griest and Seckel \cite{GriestSeckel} have pointed out when
next-to-the-lightest
superparticles (NSPs)  are slightly heavier than the LSP, this naive
treatment fails to give a correct answer. Namely, the abundance of the NSPs
is comparable to that of the LSPs around the freeze-out epoch
so that  we have to take account of
annihilation processes involving the NSPs, which
they have called {\it coannihilations}. In ref. \cite{GriestSeckel}, they
have considered a rather
accidental case where squarks are degenerate in mass with the LSP,
and showed that the relic abundance of the LSPs is greatly reduced due to
the coannihilations of the squarks. We should stress, however, the mass
degeneracy rather naturally occurs for a Higgsino-dominant LSP
\cite{MOS,Yamaguchi}.
This is because if the electroweak gauge symmetry were not broken, all
Higgsino states would be degenerate in mass. In the broken phase, the mass
splitting comes from the mixing with gauginos, which are small in the
Higgsino-dominant LSP region. Therefore the coannihilation effects will
be important in the region.

In this paper, we will  evaluate the relic abundance of the neutralino LSPs
taking account of the coannihilation processes.
We consider only the processes where neutralinos and charginos pair-annihilate
into a fermion pair.
  Since their
effects are particularly important for the LSP weighing less than the W-boson,
 we will concentrate on this case.
In a region where the Higgsino component dominates the LSP, coannihilations of
the LSP and the NSPs can be mediated by Z- and W-boson exchange. We will show
in this region, the coannihilations drastically change the previous
conclusions:
the mass density of them becomes very small, typically $\Omega_{\chi}
 h^2$ being much less than $10^{-2}$.

 The mass matrix for the four neutralinos is given by
\begin{eqnarray}
  \left( \matrix{
     M_1 & 0 & -m_Z \sin \theta_W \cos \beta & m_Z \sin \theta_W \sin \beta
    \cr
    0 & M_2 & m_Z \cos \theta_W \cos \beta & -m_Z \cos \theta_W \sin \beta
    \cr
    -m_Z \sin \theta_W \cos \beta & m_Z \cos \theta_W \cos \beta & 0& -\mu
    \cr
    m_Z \sin \theta_W \sin \beta & -m_Z \cos \theta_W \sin \beta & -\mu & 0
    \cr}
    \right)
\end{eqnarray}
in the $(\tilde B, \tilde W_3, \tilde H_1, \tilde H_2)^T $ basis.
 Here
$M_1$ and $M_2$ are the mass parameters of $\tilde B$ and $\tilde W_3$,
respectively, $\mu$ is the supersymmetric Higgs mass parameter
 and $\tan \beta$
represents the ratio of the expectation values of the two
Higgs bosons \cite{GunionHaber}.
As is done in most of the literature, we will impose the GUT
relation for the gaugino mass parameters
\begin{equation}
 M_1=\frac{5}{3} \tan^2 \theta_W M_2 \approx 0.51 M_2,
 \label{GUTrel}
\end{equation}
which is derived by solving one-loop
renormalization-group equations with $M_1=M_2$
at the GUT scale. As a phase convention, we take $M_2>0$ and $\mu$ both
positive and negative.
It is  straightforward to diagonalize the mass matrix.  The lightest of the
neutralinos denoted by $\chi_1$, which is also assumed to be the LSP, is
\begin{eqnarray}
  &  & \chi_1 = Z_{11} \tilde B+Z_{12} \tilde W_3 + Z_{13} \tilde H_1
           +Z_{14} \tilde H_2,   \cr
   & & \sum_{i=1}^{4}  Z_{1i}^{  \   2} =1.
\end{eqnarray}
Using $Z_{1i}$, we define the Higgsino purity  as
\begin{equation}
     p=Z_{13}^{\ 2} + Z_{14}^{\ 2}.
 \label{purity}
\end{equation}
When $p$ is nearly one, say $p \gsim 0.99$,
$\chi_1$ is almost a pure Higgsino, which is realized
if $M_2 \gg |\mu|$. If $M_2 \ll |\mu|$ the LSP is gaugino-like,
whilst for  $M_2 \approx |\mu| \approx m_Z$ it is a general mixture of the
four neutralinos.

  In fig. 1, we have plotted the Higgsino purity of the LSP,
$p$, defined in Eq.~(\ref{purity}) with $\tan \beta =2$ fixed.
In this paper, we will call the LSP  (almost) a pure Higgsino if
$p \gsim 0.99$, and a Higgsino-dominant LSP if $p \gsim 0.9$. The region where
$0.99 \gsim p \gsim 0.9$ will be called the Higgsino-dominant mixed region.
Fig. 2 shows the ratio of the mass difference between the LSP and the
NSP (which is a chargino or a neutralino) to the LSP mass:
$\Delta =(m_{\chi_2}-m_{\chi_1})/m_{\chi_1}$ where $m_{\chi_1}$ and
$m_{\chi_2}$ are the masses of the LSP and NSP, respectively.
We can see  the severe mass-degeneracy in the pure Higgsino LSP region.
In most of the parameter space, the NSP is the chargino.
This is also observed when we take a different value for $\tan \beta$.

In calculating the relic density including the effects of the
coannihilations, we  use the method developed by Griest and Seckel
\cite{GriestSeckel}.
Here we summarize it briefly. Let
$\chi_i$ ($i=1, \cdots, N$)\footnote{ Note that a particle and
its anti-particle
should be counted separately when they constitute a Dirac fermion.}
be  superparticles with mass $m_{\chi_i}$
($m_{\chi_1}<\cdots <m_{\chi_N}$)
 and suppose that they are nearly degenerate
in mass with the LSP, $\chi_1$.  Due to the mass degeneracy the number density
 $n_i$ of the $i$-th
particle ($i>1$), which eventually decays to the LSP, is comparable with $n_1$
around the freeze-out epoch.
The Boltzmann equation for the total of the number densities $n=\sum_i n_i$
is written
\begin{equation}
   \frac{dn}{dt}=-3Hn-\sum_{i,j} \langle \sigma_{ij}v_{rel} \rangle
                      (n_i n_j -n^{eq}_i n^{eq}_j),
\end{equation}
where
$\sigma_{ij}$ is the pair annihilation cross section of the particles $\chi_i$
and $\chi_j$, $v_{rel}$ is their relative velocity, $\langle \cdots \rangle$
denotes
the thermal average and $n_i^{eq}$ is the number density of $\chi_i$ in
thermal equilibrium. Since the reactions which
interchange the superparticles $\chi_{i}$'s with each other
occur much more rapidly than their annihilations, the ratio of $\chi_i$
density $n_i$ to the total density $n$ is well approximated by its
equilibrium value:
$n_i/n \approx n_i^{eq}/n^{eq}$. This greatly simplifies the Boltzmann
equation as
\begin{equation}
    \frac{dn}{dt}=-3Hn-\langle \sigma_{eff} v_{rel} \rangle
         (n^2-(n^{eq})^2).
    \label{Boltzmann}
\end{equation}
Therefore, we can solve eq.~(\ref{Boltzmann}) by the standard method
\cite{KolbTurner} using the {\it effective} cross section defined by
\begin{equation}
    \sigma_{eff} =\sum_{i,j} \sigma_{ij} r_i r_j,
\end{equation}
where $r_i$ represents the Boltzmann suppression of the density of the
heavier particle $\chi_i$. Explicitly
\begin{eqnarray}
     r_i& =& \frac{n_i^{eq}}{n^{eq}}=
          \frac{g_i (1+\Delta_i)^{3/2} e^{-\Delta_i x}}{\sum_{j}
            g_j (1+\Delta_j)^{3/2} e^{-\Delta_j x}}
          \propto e^{-\Delta_i x},  \\
\label{suppression}
  \Delta_i &=& (m_{\chi_i}-m_{\chi_1})/m_{\chi_1},
\label{degeneracy}
\end{eqnarray}
where $g_i$  is the  degree of freedom of the $\chi_i$ and  $x=m_{\chi_1}/T$
with $T$ being the photon temperature.

The relic abundance of the LSPs at the present day can be calculated as
\begin{equation}
    \Omega_{\chi} h^2 = \frac{1.07 \times 10^9 {\rm GeV}^{-1}}
               {g_*^{1/2} m_{Pl} \int_{x_f}^{\infty}
                \langle \sigma_{eff} v_{rel} \rangle x^{-2} d x},
 \label{abundance}
\end{equation}
where $x_f=m_{\chi_1}/T_{f}$, $T_f$ is the freeze-out temperature, $g_*$ is
the effective degree of freedom at the freeze-out epoch \cite{KolbTurner} and
$m_{Pl}$ denotes the Planck mass, $1.22 \times 10^{19}$ (GeV).
In eq.~(\ref{abundance}), we can see the relic density is roughly proportional
to the inverse of the effective cross section.

The LSPs freeze out at $x_f \sim 20$, so when $\Delta_i$ ($i>1$) is less
than about 0.1 the coannihilation effects are in general important.
When the magnitudes of the annihilation cross sections involving the heavier
particles are similar to that  of the LSP pair, the coannihilation effects
change the
relic density at most by several factors. On the other hand, if some cross
sections with the heavier particles are much larger than that of the
LSP pair, the relic abundance can be reduced by several orders of magnitude.
It will turn out that this is the case when the LSP is Higgsino-dominant.

Before giving our numerical result on the relic density,
it is illustrative to give a crude
estimate for the effects of the coannihilations in the degenerate limit
$M_2 \gg |\mu|$ (see eq. (1)).
In this limit, the lightest neutralino is nearly
either of the following states:
\begin{equation}
        \tilde H_{S,A} =\frac{1}{\sqrt{2}}(\tilde H_1 \pm \tilde H_2).
\end{equation}
The coupling of (neutralino)-(neutralino)-(Z-boson) is given
\begin{eqnarray}
   &&  \frac{1}{4}(g/\cos \theta_W) Z_{\mu}
               (\bar{\tilde H}_1 \gamma^{\mu} \gamma_5 \tilde H_1
            -\bar{\tilde H}_2 \gamma^{\mu} \gamma_5 \tilde H_2)
\cr
  & &=  \frac{1}{2}(g/\cos \theta_W) Z_{\mu} \bar{\tilde H}_A \gamma^{\mu}
                                    \gamma_5 \tilde H_S,
\label{nnZ}
\end{eqnarray}
while that of (neutralino)-(chargino)-(W-boson) is
\begin{equation}
    \frac{g}{2} W_{\mu}^-
    (\bar{\tilde H}_A \gamma^{\mu} \tilde H^+
    +\bar{\tilde H}_S \gamma^{\mu} \gamma_5 \tilde H^+) +h.c.,
\end{equation}
where $g$ is the $SU(2)_L$ gauge coupling constant.
 Eq.~(\ref{nnZ}) implies that in a region of the Higgsino-dominant LSP, the
coupling of the LSPs to the Z-boson is very suppressed: it is proportional to
$Z_{13}^{\ 2}-Z_{14}^{\ 2} \sim m_Z^2/(M_2 \mu)$, which vanishes at $M_2
\rightarrow \infty$ limit.

The annihilation of the LSP pair to fermions is dominated by this suppressed
Z-boson exchange process, since the couplings of the other processes,
{\it i.e.} the Higgs boson exchange and the sfermion exchange, are even
smaller.  Moreover the $s$-wave annihilation
to a  fermion pair is suppressed as the initial
state is a pair of the identical Majorana fermions. Therefore the authors of
the previous papers have
concluded that the Higgsino-dominant LSPs which are lighter
than the W-boson are (too)  rich in the Universe.

The reality is, however, they disappear through the coannihilations.
To see the coannihilation effects,
 consider the coannihilation process $\tilde H_{S,A}
\tilde H^{\pm}  \rightarrow$ (fermions) mediated
by the W-boson exchange in the $s$-channel. The cross section for this
process is much larger than that of the annihilation of the LSP pair
discussed above, because
there is no suppression factor such as $Z_{13}^{\ 2}-Z_{14}^{\ 2} $ in the
couplings to the weak boson and the annihilation
occurs  in the $s$-wave. Indeed it  is estimated as
\begin{equation}
  \langle \sigma v_{rel} \rangle \sim
    \frac{9 g^4}{16 \pi m_W^2} f(\mu)
  \sim
    5 f(\mu) \times 10^{-6} {\rm GeV}^{-2}
\end{equation}
with $f(\mu)=(\mu/m_W)^2/(4(\mu/m_W)^2-1)^2$.
The contribution to the effective cross section
is therefore given by
\begin{equation}
     \langle \sigma v_{rel} \rangle r_C
   \sim 5 f(\mu) r_C \times 10^{-6} {\rm GeV}^{-2},
 \label{crsc}
\end{equation}
where $r_C \sim e^{-\Delta_C x_f} \sim e^{-20 \Delta_C}$.
$\Delta_C$ and $r_C$ represent the mass degeneracy and the
Boltzmann suppression factor for the chargino, respectively
(see eqs.~(\ref{degeneracy}) and (8)).
A quite similar estimate can be obtained for the cross section with the
second lightest neutralino, which is in the same  order of the magnitude.
Since, as we mentioned above, the chargino is the NSP in a large part of the
parameter space, the effective cross section is dominated by eq.~(\ref{crsc}),
which yields the estimate of the relic abundance
\begin{equation}
   \Omega_{\chi} h^2 \sim \frac{5 \times 10^{-10} {\rm GeV}^{-2}}
                              {\langle \sigma_{eff} v_{rel} \rangle}
                      \sim \frac{10^{-4}}{r_C f(\mu)}.
\end{equation}
Therefore, we can expect that for $\Delta_C \lsim 0.2$ ({\it i.e.}
$r_C \gsim 10^{-2})$, the relic abundance of the LSPs will become very small
due to the coannihilations.

It is tedious but straightforward to calculate the effective cross section
accurately for a general set of the parameters ($\mu$, $M_2$).
 In fig. 3, we show  the abundance of the neutralino LSPs,
including the coannihilation processes.
We have chosen $\tan \beta=2$,  all squark and slepton masses equal to
1 TeV and the pseudoscalar mass $m_A=1$ TeV.
Since the coannihilations occur mainly through the Z- or W-exchange, their
effects do not depend on the choice of the masses of the sfermions and
the pseudoscalar. We can obtain qualitatively a similar plot for a different
value of $\tan \beta$.
For comparison, we have given in fig. 4, the same plot but without
taking account of the coannihilation processes.
 By comparing the two figures, the importance of the coannihilations is
manifest. In the pure Higgsino region with the purity $p \gsim 0.99$, the mass
degeneracy is severe, {\it i.e.} $\Delta \lsim 0.1 $, and
the coannihilations greatly reduce the neutralino density. In most of the
region, we can see  $\Omega_{\chi} h^2 \ll 10^{-2}$.
Thus the LSPs  cannot constitute any form of the dark matter,
{\it i.e.} not only the dark matter of the universe but also the dark matter
of the galactic halos.
This contrasts with the conclusion obtained previously
where the density of the light
Higgsinos can reach the critical one. Even for the Higgsino dominant-mixed
region  ($0.9 \lsim p \lsim 0.99$) where the mass degeneracy is not so severe,
the coannihilation effects
are significant.  They reduce the density
with an order of magnitude or more.

In this paper,
 we have considered  the coannihilation effects to calculate
the present abundance of the neutralino LSPs whose mass is less than that of
the W-boson. We have shown that in
 the Higgsino-dominant LSP region the
relic density gets very small. On the other hand, as mentioned above
for a  Higgsino-dominant LSP heavier than the W-boson,
the annihilation to the W-pair is kinematically allowed
and $\Omega_{\chi} h^2$ is much smaller than order unity until the LSP mass
reaches the TeV region.
Combining these two results,
 we can conclude that the Higgsino-dominant
 LSP is no longer an interesting candidate
for the  dark matter.

We would like to make two comments:
firstly our calculation does not contain annihilation processes whose final
state is other than a fermion pair. When Higgs bosons  are light, the LSPs and
NSPs will annihilate to them significantly. If this is the  case, the
relic density becomes even smaller.  Moreover
we may consider a process such as
\begin{equation}
   \tilde H_{S,A} \tilde H^{\pm}
    \rightarrow W^{\pm} \rightarrow \gamma W^{\pm},
\end{equation}
which would enhance the effective cross section by  several factors.
Although it will  not change our result drastically, a more detailed and
accurate analysis is significant \cite{MizutaYamaguchi}.
Secondly there are other cases where the
 coannihilations are important. For example,
when the GUT relation for the gaugino masses (\ref{GUTrel}) is relaxed,
the neutral Wino
 can become the LSP \cite{Yamaguchi}. Since it is highly
degenerate in mass with its charged counterparts, the coannihilations
involving
them dominate the effective annihilation cross section and hence their
relic abundance is very small. This issue will
be discussed in a separate publication \cite{MizutaNgYamaguchi}.

We would like to thank T. Yanagida for careful reading of the manuscript and
useful comments. We are also grateful to H. Murayama and T. Yanagida for
discussions.

\newpage


\section*{Figure captions}

\newcounter{Figures}

\begin{list}
{{\bf Fig. \arabic{Figures}.}}
{\usecounter{Figures}}
\item
The Higgsino purity of the LSP, $p$ ,defined by $p=Z_{13}^2+Z_{14}^2$;
(a)$\mu > 0$ and (b)$\mu < 0$.
We have chosen $\tan \beta =2$.
\lq \lq LEP" means the excluded region by LEP constraints
and \lq \lq $M_{\chi}>M_W$" means the region where the lightest neutralino
is heavier than W-boson with which we are not concerned.
\label{fpurity}
\item
The ratio of the mass difference between the LSP and the NSP to
the LSP mass, $\Delta$ ,defined by
$\Delta = (m_{\chi_2}-m_{\chi _1})/m_{\chi _1}$;
(a) $\mu > 0$ and (b) $\mu < 0$.
We have chosen $\tan \beta =2$.
The meanings of \lq \lq LEP" and \lq \lq $M_{\chi}>M_W$" are
the same as in fig. \ref{fpurity}.
\label{fdegeneracy}
\item
The relic abundance of the neutralino LSPs,
including the coannihilation processes;
(a) $\mu > 0$ and (b) $\mu < 0$.
We have chosen $\tan \beta=2$, all squark and slepton masses equal to 1TeV
and the pseudoscalar mass $m_A=1$TeV.
Note that the coannihilation effects are independent of
the squark and slepton masses and the pseudoscalar mass.
The meanings of \lq \lq LEP" and \lq \lq $M_{\chi}>M_W$" are
the same as in fig. \ref{fpurity}. By comparing this with fig. 4, we can find
that the coannihilations greatly reduce the relic abundance of the LSPs in
the Higgsino-dominant LSP region ($M_2 \gg |\mu|$).
\label{relic-co}
\item
Same as fig. \ref{relic-co} but without taking account of the coannihilation
processes.
\label{relic-LSP}
\end{list}

\end{document}